\documentclass[aps,showpacs,prl,twocolumn]{revtex4}

\usepackage{graphicx}
\usepackage{amsmath}
\usepackage[usenames,dvipsnames]{color}

\newcommand{\bra}[1]{\left\langle #1\right|}
\newcommand{\ket}[1]{\left|#1\right\rangle}

\begin{document}
\title{Spatial Patterns of Rydberg Excitations from Logarithmic Pair Interactions}

\author{Wolfgang Lechner}
\email{w.lechner@uibk.ac.at}
\author{Peter Zoller}

\affiliation{Institute for Quantum Optics and
Quantum Information, Austrian Academy of Sciences, 6020 Innsbruck, Austria}
\affiliation{Institute for Theoretical Physics,  University of Innsbruck,
6020 Innsbruck, Austria}

\date{\today}


\begin{abstract} 
The collective excitations in ensembles of dissipative, laser driven ultracold atoms exhibit crystal-like patterns, a many-body effect of the Rydberg blockade mechanism. These crystalline structure are revealed in experiment from a post-selection of configurations with fixed numbers of excitations. Here, we show that these sub-ensemble can be well represented by ensembles of effective particles that interact via logarithmic pair potentials. This allows one to study the emergent patterns with a small number of effective particles to determine the phases of Rydberg crystals and to systematically study contributions from $N$-body terms.  
\end{abstract}
\maketitle

Current experiments with laser driven, trapped ultracold Rydberg atoms \cite{RYDBERGCYRSTAL,URBANEXP,PFAUEXP,WEIMERNATURE,GRANGIER,WEIDEMULLER} open a viable route to study driven dissipative many-body systems in the strongly correlated regime \cite{RYDBERGBOOK,SAFFMANREV,JAKSCH,POHL,ZWERGER,WEIMER,BLOCKADE,TONG}. In these experiments atoms are prepared in the ground state and are collectively driven by a laser to a Rydberg state. Due to dipole-dipole or van-der Waals interactions, excitations of adjacent particles are suppressed, the socalled Rydberg blockade mechanism [Fig. \ref{fig:illustration}]. In ongoing experiments, the position of individual Rydberg excitations can be determined in single-shot single-particle resolved measurements \cite{RYDBERGCYRSTAL} [a typical configuration is illustrated in Fig. \ref{fig:illustration}(a)]. These excitation configurations are then post-selected into sub-ensembles according to the number of excitations $N$, resulting in a conditional density $\rho(\mathbf{x}|N)$, where $\mathbf{x}$ represents the positions of all $N$ excitations [see Fig \ref{fig:illustration}(b)]. It is these sub-ensembles that exhibit the remarkable crystal-like density patterns known as Rydberg crystals. While the short time coherent regime has been studied experimentally, we will focus here on the steady state including dissipation \cite{PETROSYAN,PETROSYAN2}. As the blockade radius may be large compared to the the lattice spacing, the number of atoms $N_a$ can be several orders of magnitude larger than the number of excitations $N$.  

\begin{figure}[ht]
\centerline{\includegraphics[width=8cm]{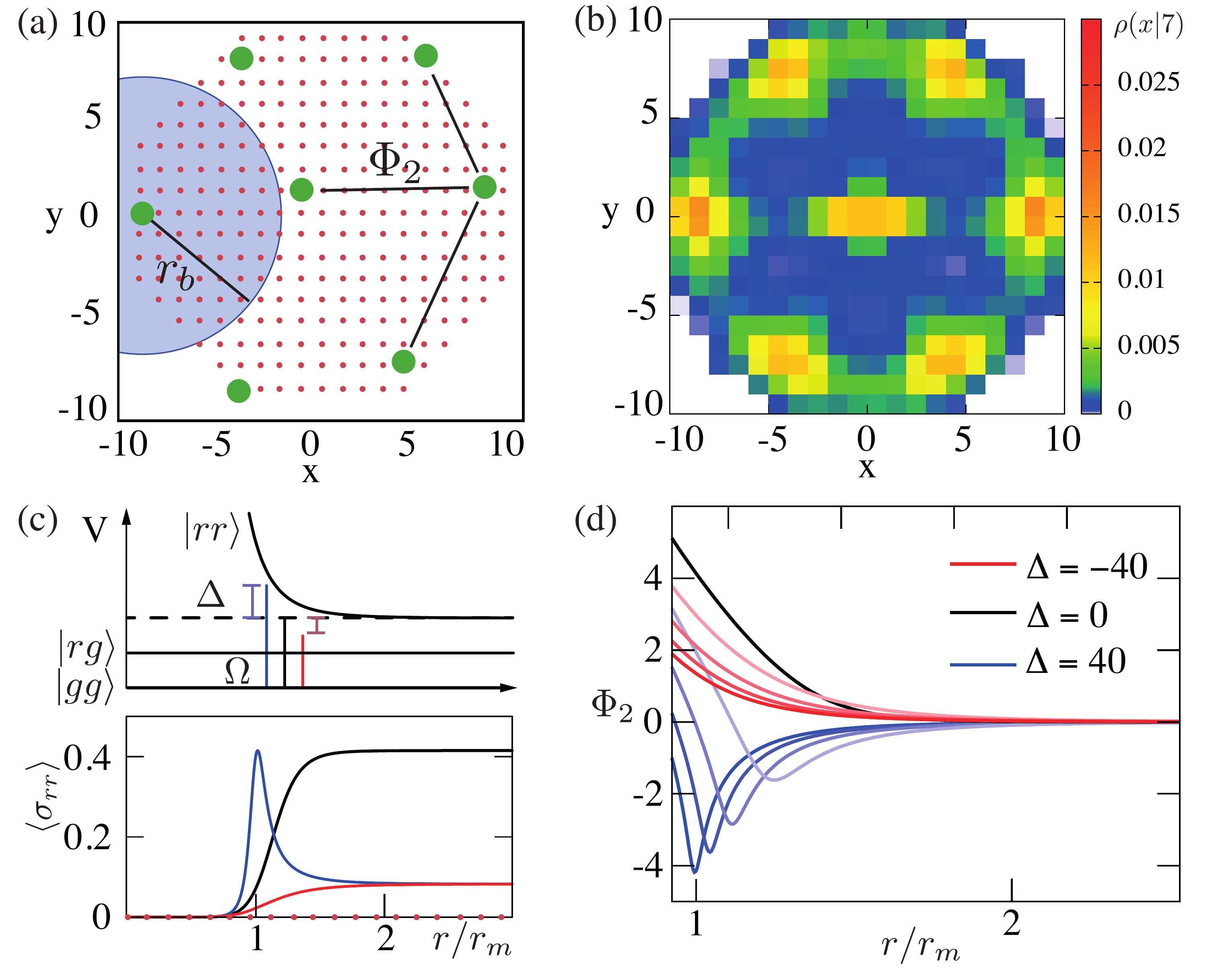}}
\caption{(a) The system of interest consists of ground-state atoms (red dots) which are laser excited to Rydberg states (green dots) in an optical lattice (see e.g. Ref. \cite{RYDBERGCYRSTAL}). Excitations within a blockade radius $r_b$ are suppressed. (b) Conditional density $\rho(x|N=7)$ from the logarithmic potential model. (c) Excitation probability $\langle \sigma_{rr} \rangle$ (see main text) as a function of the distance $r$ in units of the excitation maximum $r_m = (\Delta/C_6)^{1/6}$ for $\Delta = 40$. For resonant ($\Delta = 0$, black) and red-detuned driving $\Delta < 0$ (red) excitations are suppressed at short distances. When driving with a blue-detuned laser $\Delta > 0$  (blue) the excitation probability is peaked at a resonant distance $r_m$. (d) Logarithmic potential Eq. \eqref{equ:V} as a function of the distance $r$ between two effective particles. In the case $\Delta = 0,-40,-30,-20,-10$ (black and red), the effective particles are purely repulsive. For blue detunings $\Delta = 10,20,30,40$ (blue) the excitation resonance translates into a potential energy minimum. (Other parameters in units of $\Omega$ are: $C_6=1$,$\Gamma_R=0.1$, $\Gamma_Z=1$).  }
\label{fig:illustration}
\end{figure}

In this Letter we present an analytic effective particle model for excitations in dissipative Rydberg crystals. This model is derived from the conditional density, which can be written as a sum of $N$-body terms 
\begin{eqnarray}
& & \rho(\mathbf{x}|N) = \exp[\Phi(\mathbf{x})] = \label{equ:rep} \\ \nonumber 
&=& \exp[\Phi_0 + \sum_{i} \Phi_1(x_i) + \sum_{ij} \Phi_2(x_i,x_j) + \\ \nonumber
&+& \sum_{ijk} \Phi_3(x_i,x_j,x_k) + ... ].
\end{eqnarray}
Here, the first term $\Phi_0$ is a global shift, $\Phi_1(x_i)$ represents a term that only depends on single particle positions, $\Phi_2(x_i,x_j)$ is a pair correlation term, etc. Below we will present an analytic expression for $\Phi_2$ with a  logarithmic distance dependence. We refer to this term as \textit{logarithmic pair interaction} in analogy to particle interactions in statistical mechanics [see Fig. \ref{fig:illustration}(d)]. We show that the steady state of Rydberg crystals $\rho(\mathbf{x}|N)$ is well described by the interaction $\Phi_2$ for all $N$. In particular, for van-der Waals interactions it is the dominant term. We find that higher order contributions show a spatial pattern which indicate that these terms soften the crystals (see Fig. \ref{fig:rydbergcrystal}). Results are compared with the steady state solution of the quantum master equation [Eq. \ref{eq:qme}] including quantum correlations for small systems and with a Monte Carlo scheme from Ref. \cite{PETROSYAN} for large system sizes, respectively. 
 
\begin{figure}[ht]
\centerline{\includegraphics[width=8.5cm]{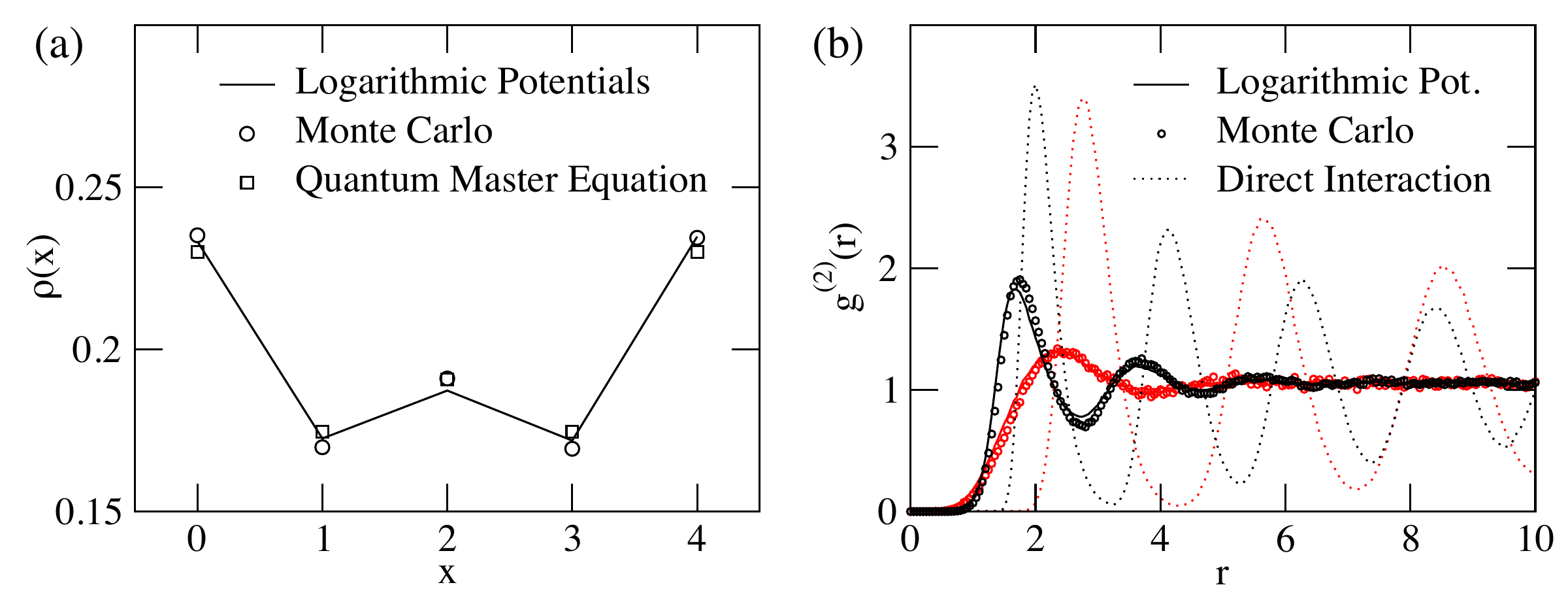}}
\caption{(a) Steady-sate density $\rho(x)$ of excitations from the quantum master equation [Eq. \ref{eq:qme}] (squares) compared to the results of the Monte Carlo scheme from Ref. \cite{PETROSYAN} (circles) and the logarithmic potentials (solid line). Here, $N_a=5$, $\Omega = 1$, $\Delta = 0$, $\Gamma_z/\Gamma_r = 50$ and $r_b=1$ with a van-der Waals direct interaction $V(r) \propto 1/r^6$. (b) Pair correlation function of excitations from logarithmic potentials (solid line) and Monte Carlo (circles). For both, van-der Waals interaction $V(r) \propto 1/r^6$ (black) and dipole-dipole interaction $V(r) \propto 1/r^3$ (red), the results are in excellent agreement. For comparison, a crystal that results from the direct interaction $1/r^6$ or $1/r^3$, respectively, has completely different features (dotted). The parameters are chosen as in Ref. \cite{PETROSYAN}:  $\Omega = 1$, $\Delta = 0$, $\Gamma_r=0.1 \Omega$, $\Gamma_z=\Omega$. The blockade radius was set to $r_b=20 a_0$. }
\label{fig:fig2}
\end{figure}

\textit{Model} - We consider a laser driven system of ultracold trapped atoms described by the Hamiltonian 
\begin{equation}
H = \sum_i^{N_a} \Delta \sigma_{rr}^{(i)} + \sum_i^{N_a} \frac{\hat{\Omega}}{2} (\sigma_{gr}^{(i)} + \sigma_{gr}^{(i)}) + \sum_{i \neq j } V_{ij}  \sigma_{rr}^{(i)}  \sigma_{rr}^{(j)}.
\label{equ:H}
\end{equation}
Here, $N_a$ is the number of atoms, $\Delta$ is the detuning, $\sigma_{xy}^{(i)} = \ket{x} \bra{y}$ are the matrix elements coupling states $x,y \in \{ r,g \}$, where $g$ is the ground state and $r$ is the Rydberg-state, evaluated for particle $i$ and the bare Rabi frequency is $\hat{\Omega}$. The interaction is $V_{ij}(r)=C_6/r_{ij}^6$ (van-der Waals) or $V_{ij}(r)=C_3/r_{ij}^3$ (dipole-dipole)  with $r_{ij}$ the distance between atom $i$ and $j$. In the following energies are given in units of $\Omega$ and distances in units of $a_0$, the lattice spacing between atoms. We are interested in the steady state in the open dissipative regime described by the master equation 
\begin{equation}
\frac{\partial \rho}{\partial t} = -\frac{i}{\hbar} [H,\rho] + \sum_{j=0}^{N_a} (\mathcal{L}_r^j \rho + \mathcal{L}_z^j \rho )
\label{eq:qme}
\end{equation}
with $\mathcal{L}_r^j \rho = \frac{1}{2} \Gamma_r [2 \sigma_{rg}^j\rho\sigma_{rg}^j - \{\sigma_{rr}^j ,\rho \}]$ and  $\mathcal{L}_z^j \rho = \Gamma_z [(\sigma_{rr}^j - \sigma_{gg}^j)\rho(\sigma_{rr}^j - \sigma_{gg}^j) - \rho ]$ (details see \cite{PETROSYANBOOK}). In a typical experiment, the decay rate is small $\Gamma_r \ll \Omega$. Dephasing results from various mechanisms such as finite linewidth of the excitation laser and particle motion which can be of the order of $\Gamma_z \approx \Omega$. The steady state of the quantum master equation can be calculated numerically exact for small numbers of atoms (typically $N_a < 20$). An example of $\rho(x)$ is shown in Fig. \ref{fig:fig2} (a) for $N_a=5$. The conditional density is then deduced from tracing out all configurations with excitations 
$\rho(\mathbf{x}|N) = \mathrm{Tr}[ \rho \delta(\hat{N} - N)]$, where $\hat{N} = \sum_i^{N_a} \langle \sigma_{rr}^{(i)} \rangle$ and $\delta$ is the Dirac delta function.

\begin{figure}[ht]
\centerline{\includegraphics[width=8cm]{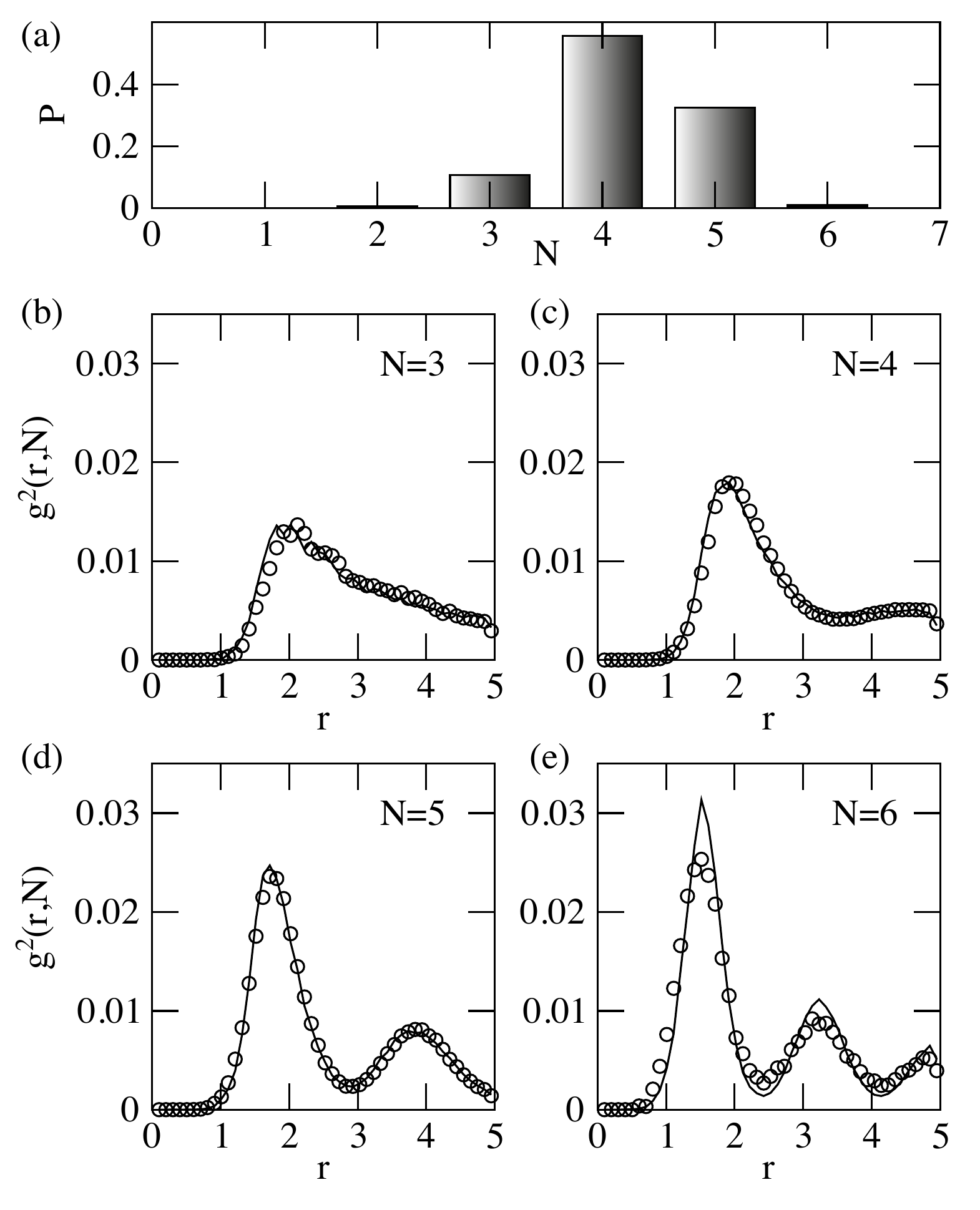}}
\caption{(a) Distribution of excitations in a 1D chain of $N_a=200$ atoms, $\Delta=0$, $C_6 = 10.0 \Omega$, $\Gamma_r=0.01 \Omega$, $\Gamma_z=0.1 \Omega$. The average number of excitations is $\hat{N}=4.4$ and configurations with $N=3,4,5,6$ are observed. (b-e) The density-density pair correlation function $g_2(r,N)$ (dots) is almost identical to the result of the logarithmic potential model (solid) for each sub-set with $N$ particles. }
\label{fig:comparison}
\end{figure}

\textit{Logarithmic Pair Interaction -}  In the following we systematically study the expansion Eq. (\ref{equ:rep}). As the simplest approximation, the constant term $\Phi_0$ and all higher order terms $\Phi_3,\Phi_4,...$ are neglected. In this case, the only remaining term is the pair potential $\Phi_2$. Note, that an external trapping potential can be added using the $\Phi_1$ term. Further we assume that the term $\Phi_2 (x_i,x_j) = \Phi(|x_i-x_j|)$ is a function of the distance only. In this approximation the steady state probability of finding a configuration $\mathbf{x}$ in the ensemble with $N$ excitations is

\begin{equation}
P_N(\mathbf{x}) = \frac{1}{Z_N} \exp[\sum_{i<j}^N \Phi_2(x_i-x_j)].
\label{equ:pphi2}
\end{equation}
Here, the probability is normalized with the the partition sum $Z_N$. In this approximation, the probability of states Eq.~\eqref{equ:pphi2} has the form of a canonical distribution. In the following we derive a analytic expression for the potential $\Phi_2$. 

For two particles, it has been shown \cite{PETROSYAN} that in the large time limit $t \gg \Gamma_r,\Gamma_z $ the steady state probability of two atoms at distance $r$ is ${\langle \rho_{rr} \rangle = \Omega^2/\{2\Omega^2 + \frac{\Gamma}{\gamma_{rg}}(\gamma_{rg}^2 + [\Delta + V(r)]^2)\}}$. Here, $\gamma_{rg} = \frac{1}{2}\Gamma_r + 2 \Gamma_z$ with $\Gamma_r$ the decay rate, $\Gamma_z$ the dephasing and $\Omega$ the $\sqrt{N_r}$-enhanced Rabi frequency, where $N_r$ is the number of atoms within a blockade radius \cite{BLOCKADE, PETROSYAN}. In the case of van der Waals interactions, $V(r) = C_6/r_{ij}^6$ if $i$ and $j$ are excited and $V(r)=0$ otherwise illustrated in Fig. \ref{fig:illustration}(a) together with the excitation probability as a function of the distance $\langle \rho_{rr} \rangle$. The length-scale of the a superatom is given by the blockade distance which is defined for van-der Waals interactions as $r_b = (C_6/(2 \Omega)\sqrt{\Gamma_R/\gamma_{rg}})^{1/6}$ \cite{PETROSYAN}. Applying the reversible work theorem \cite{CHANDLER} to connect the pair correlation with a pair interaction we find $\Phi_2(r_{ij}) = -\ln[\sigma_{rr}(r)] + \ln[\sigma_{rr}(\infty)]$ which can be written as 
\begin{equation}
\Phi_2(r_{ij}) = \ln \left[ 1+\frac{ (\frac{C_6}{r_{ij}^6})^2+ 2 \Delta \frac{C_6}{r_{ij}^6} }{ \frac{2 \gamma_{rg}}{\Gamma}  \Omega^2 + \gamma_{rg}^2 + \Delta^2 } \right] . 
\label{equ:V}
\end{equation}
Here, $r_{ij}=x_i - x_j$ is the distance between the effective particle $i$ and $j$.  Examples of this potential with detunings in the range $\Delta = [-40,40]$ are depicted in Fig. \ref{fig:illustration}(d). This particle model allows one to interpret the patterns of post-selected Rydberg excitations as the result of $N$-particles with a logarithmic pair interaction. In particular by interpreting $\Phi_2 = V_{\textrm{eff}}/(k_B T)$, where $T$ is the temperature, $k_B$ is the Boltzmann constant, and $V_{\textrm{eff}}$ the effective potential energy of the particles, the system follows a canonical ensemble. For a comparison of thermal fluctuations and fluctuations in the microscopic picture see \cite{suppmat}.  Note, that in this interpretation the temperature is not a tunable parameter. We note, that for blue detuning, the first two orders of a series expansion at the minimum are similar to a Lennard-Jones potential. 

\textit{Numerical Results} - For small system sizes  Eq. (\ref{eq:qme}) can be solved numerically and compared to the logarithmic potential model. For large system sizes this is not feasible and we compare our results with results from a Monte Carlo scheme introduced in Ref. \cite{PETROSYAN} which is shortly summarize here for completeness. In this scheme each configuration is represented $N_a$ atoms where $N$ are excited. In each Monte Carlo step, transitions between configurations from $N$ excitations to either $N-1$ or $N+1$ excitations are possible with the following protocol: a random atom $i$ is chosen and, given that it is in the ground state, it is excited with an acceptance probability $P^{(i)}_{acc} = \Omega^2/(2\Omega^2 + \frac{\Gamma}{\gamma_{rg}}[\gamma_{rg}^2 + (\sum_{i<j} V(r_{ij})+ \Delta)^2 ])$. Accordingly, if the chosen atom is excited it is de-excited with probability $1- P_{acc}$. Note, that the logarithmic potential model acts on sub-sets of constant number of excitations, and the sequence of de-excitation and excitation is translated to a dynamics of an excitation due to pair interactions \cite{suppmat}. A useful measure to compare configurations is the pair correlation function of excitations 
\begin{equation}
g^2(r) = \frac{1}{\rho}\langle  \sum_{i \neq j} \delta(\mathbf{r} - [\mathbf{r_i} - \mathbf{r_j}]) \rangle,
\label{equ:g2}
\end{equation}
where angular brackets $\langle \cdot \rangle$ denote the ensemble average in the the canonical ensemble and the average over all configurations. 

The comparison for small system sizes as depicted in Fig. \ref{fig:fig2}(a) shows a good agreement between the results of the quantum master equation, the Monte Carlo procedure from Ref. \cite{PETROSYAN} and the logarithmic potential model. The density $\rho(x)$ in the logarithmic potential model is calculated from the $\rho(x) = \sum_n c_n \rho(x|n)$, where the sum goes from $n=1$ to $n=5$ and the coefficients $c_n$ is are taken from an independent Monte Carlo simulation. 
 
Fig. \ref{fig:fig2}(b) depicts the results of the pair correlation functions from Monte Carlo calculations in comparison to the logarithmic potential model for a large $1D$ system. For the chosen parameters, the average number of excitations is $N = 20$. We find that the spatial excitation statistics from the dynamics of particles with logarithmic pair interaction is almost identical to the results from the Monte Carlo simulation. For comparison, the crystal structure from particles with a direct van-der-Waals or dipole-dipole interaction exhibits a completely different pattern.

The logarithmic potential model does not just reproduce the pattern from an average over all possible excitation numbers $\rho(x)$ but also for each sub-set of ensembles $\rho(x|N)$. Figs. \ref{fig:comparison} depict the results for all  numbers of excitations that are possible for a given blockade radius. The distribution of excitations from the Monte Carlo simulation [Fig. \ref{fig:comparison}(a)] is of Gaussian shape and peaked around $N\approx4$. To compare the pair correlation function for each sub-set of constant number of excitations we consider  
\begin{equation}
g^2(r,N) = \frac{1}{\rho}\langle  \sum_{i \neq j} \delta(\mathbf{r} - [\mathbf{r_i} - \mathbf{r_j}]) \rangle_N,
\label{equ:g2}
\end{equation}
where the  $\langle A \rangle_N = \langle A \delta(\hat{N} - N)\rangle$ denotes averages with a constraint on the number of excitations $N$. The results are shown in panels Figs. \ref{fig:comparison}(b)-(e) ranging from $N=3$ to $N=6$. For small numbers of excitations the system is in a liquid-like state while for larger densities the structures are crystal like. These results from the Monte Carlo simulation are in good agreement with the results from the canonical ensemble of logarithmic pair interactions with the corresponding numbers of particles. 

\begin{figure}[ht]
\centerline{\includegraphics[width=8.5cm]{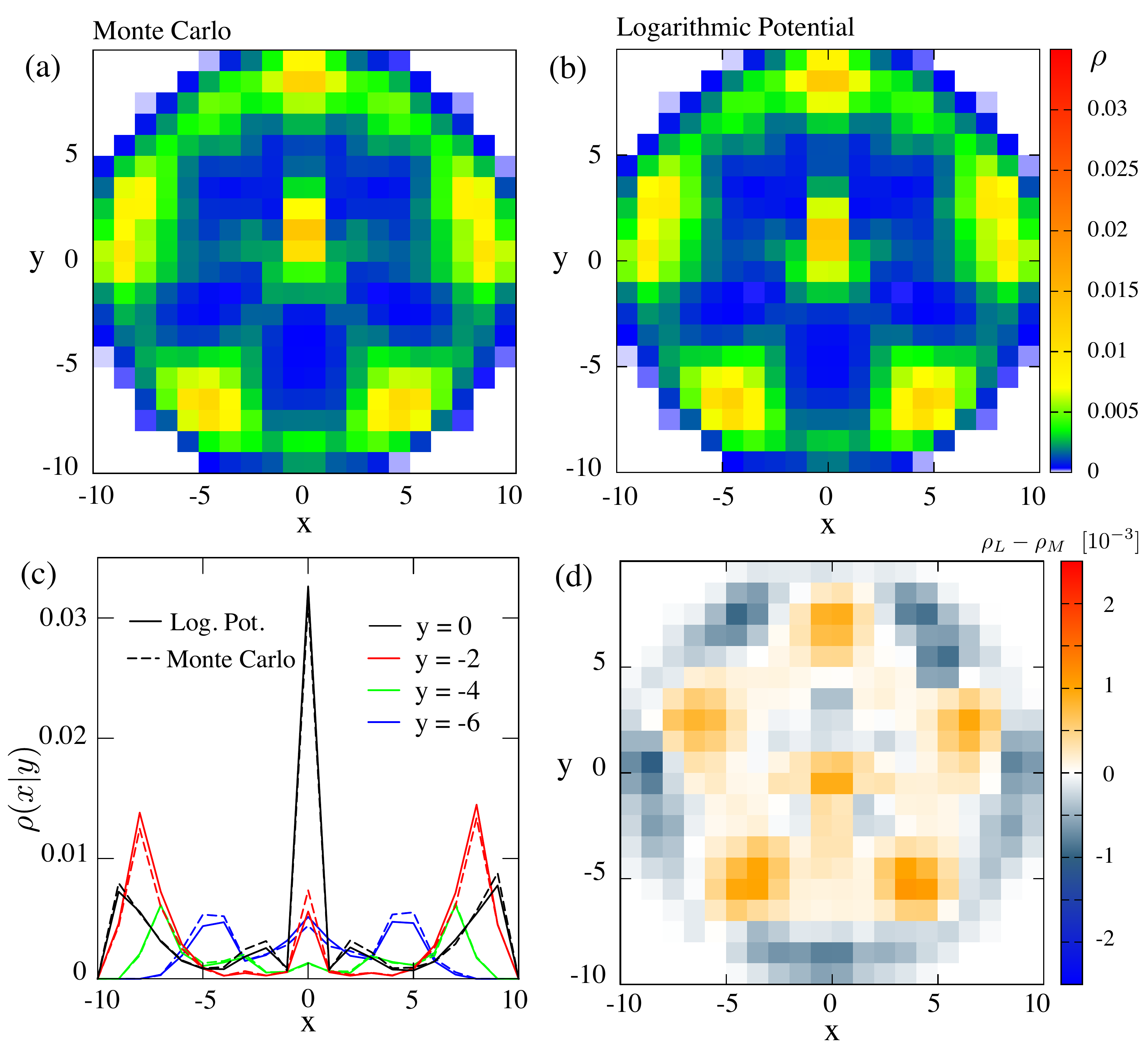}}
\caption{Comparison of the density of states after the rotation protocol from the full master equation with post selection of configurations with $N=6$ excitations (a) and from logarithmic potentials with $N=6$ particles (b). Note, that the shape is the result of a mixture of a hexagon and a pentagon from configurations with all $6$ excitations at the border and configurations with $5$ excitations on the border an an excitation in the center. (c) Comparison of the steady state probability of excited atoms in the $N=6$ manifold as a function of $x$ for fixed $y$. The results from the logarithmic potential (solid) and master equation (dashed) are in good agreement. (d) The difference in density between the two models allows one to estimate the effects of three body (as well as higher order) terms. Here, the $N$-body contributions are of the  the order of $10^{-4}$ (approximately $1\%$) and the pattern indicates that these contributions destabilize crystallinity. }
\label{fig:rydbergcrystal}
\end{figure}

As a two-dimensional example we study a disc-like setup as used in Ref. \cite{RYDBERGCYRSTAL}. In this setup, atoms are loaded into a deep optical lattice. Then, atoms with a distance from the origin larger than $r_0$ are removed (see  Fig. \ref{fig:rydbergcrystal}(a)]. The parameters are chosen such that there are approximately $N_a=20\times20$ sites and the blockade distance $r_b\approx r_0$. This allows for Rydberg crystals with $N \lesssim 8$ excitations. The system is rotationally invariant and therefore a rotational invariant averaging protocol was introduced in Ref. \cite{RYDBERGCYRSTAL}. In this protocol, described in more detail in \cite{suppmat}, each configuration is reorientated to eliminate the rotational invariance. The density of excitations after this rotation is shown for $N=6$ in Fig. \ref{fig:rydbergcrystal}(a) from Monte Carlo simulations and Fig. \ref{fig:rydbergcrystal}(b) from the logarithmic potential model. For $N=6$ two symmetries are possible: (i) either all $6$ excitations are positioned at the surface and the crystal is of hexagonal shape, or (ii) there is $1$ excitation in the center is surrounded by $5$ excitations. In the latter case the crystal has a $5$-fold rotation symmetry. This mixture of two crystalline structures is clearly seen in Figs. \ref{fig:rydbergcrystal}(a) and also reproduced with the logarithmic potentials. The results are also in quantitative good agreement as shown in Fig. \ref{fig:rydbergcrystal}(b). Comparing the densities quantitatively [see Fig. \ref{fig:rydbergcrystal}(c)] shows that for van-der Waals interactions the deviation is of the order of $1\%$. These deviations are, however, not uniform in the system. The spatial distribution [Fig. \ref{fig:rydbergcrystal}(d)] shows that three-body terms destabilize the crystalline pattern. 

In conclusion, we have presented a framework that allows one to treat a system  described by excitations and de-excitations as a series of dynamical many-body systems with logarithmic pair interactions. This allows one to predict Rydberg crystal patterns with well-established tools from statistical mechanics.A practical advantage is, that the number of variables is reduced from $N_a$ to $N$ variables which is considerable smaller for a large blockade radius. The expansion of the density into potentials with $N$-body terms also opens a way to study three body contributions. As an outlook, the mapping and the statistical mechanics description of Rydberg crystals may be interesting as an extension to previous \textit{hard rod} models of Rydberg excitations \cite{HARDRODS,HARDRODS2}. Further, the mapping from the excitations model to a interaction model may also serve as the starting point for the reverse process, where pair interactions are designed with a particular purpose (e.g. formation of a particular structure, anisotropic interaction) which can be mapped back on a excitation protocol. Finally, the multi-particle terms which are small in the case of van-der Waals interactions may become dominant in systems with long range interactions and may open a new way to design multi-particle interactions. 

\textit{Acknowledgements} - We thank I. Bloch, M. Fleischhauer, T. Pohl, I. Lesanovsky, and M. Baranov for fruitful discussions. Work was supported by the Austrian Science Fund (FWF): P 25454-N27, SFB FoQuS and ERC Synergy Grant UQUAM.

\bibliographystyle{prsty}

\begin{thebibliography}{10}
\bibitem{RYDBERGCYRSTAL}P. Schau\ss,	 M. Cheneau,	 M. Endres, T. Fukuhara, S. Hild, A. Omran, T. Pohl, C. Gross, S. Kuhr and I. Bloch, {\em Nature} {\bf 491}, 87 (2012).
\bibitem{URBANEXP}E. Urban, T. A. Johnson, T. Henage, L. Isenhower, D. D. Yavuz, T. G. Walker and M. Saffman, Nature Physics {\bf 5}, 110 (2009).
\bibitem{PFAUEXP}R. Heidemann, U. Raitzsch, V. Bendkowsky, B. Butscher, R. L\"ow, L. Santos, and T. Pfau, Phys. Rev. Lett. {\bf 99}, 163601 (2007).
\bibitem{WEIMERNATURE}H. Weimer, M. M\"uller, I. Lesanovsky, P. Zoller and H. P. B\"uchler, Nature Physics {\bf 6}, 382  (2010). 
\bibitem{GRANGIER}A. Ga\"etan, Y. Miroshnychenko, T. Wilk, A. Chotia, M. Viteau, D. Comparat, P. Pillet, A. Browaeys and P. Grangier, Nat. Phys. 5, 115 (2009).
\bibitem{WEIDEMULLER} H. Schempp, G. G\"unter, M. Robert-de-Saint-Vincent, C. S. Hofmann, D. Breyel, A. Komnik, D. W. Sch\"onleber, M. G\"arttner, J. Evers, S. Whitlock, and M. Weidem\"uller, Phys. Rev. Lett. {\bf 112}, 013002 (2014).
\bibitem{RYDBERGBOOK} T. F. Gallagher, {\it Rydberg Atoms} (Cambridge University Press, Cambridge, 1994).
\bibitem{SAFFMANREV} M. Saffman, T. Walker, and K. M\o lmer, Rev. Mod. Phys. {\bf 82}, 2313–2363 (2010).
\bibitem{JAKSCH}D. Jaksch, J. I. Cirac, P. Zoller, S. L. Rolston, R. C\^ote ,and M. D. Lukin, Phys. Rev. Lett. 85, 2208 (2000).
\bibitem{POHL}T. Pohl, E.  Demler, and M. D. Lukin, Phys. Rev. Lett. {\bf 104}, 043002 (2010).
\bibitem{ZWERGER}I. Bloch, J. Dalibard, and W. Zwerger,  Rev. Mod. Phys. {\bf 80}, 885–964 (2008)
\bibitem{WEIMER}H. Weimer, R. L\"ow, T. Pfau, and H. P. B\"uchler,  Phys. Rev. Lett. {\bf 101}, 250601 (2008).
\bibitem{BLOCKADE}M. D. Lukin, M. Fleischhauer, R. Cote, L. M. Duan, D. Jaksch, J. I. Cirac, and P. Zoller, Phys. Rev. Lett. {\bf 87}, 037901 (2001).
\bibitem{TONG}D. Tong, S. M. Farooqi, J. Stanojevic, S. Krishnan, Y. P. Zhang, R. C\^ote, E. E. Eyler, and P. L. Gould, Phys. Rev. Lett. {\bf 93}, 063001 (2004).
\bibitem{DEPHASING}A. W. Glaetzle, R. Nath, B. Zhao, G. Pupillo, and P. Zoller, Phys. Rev. A {\bf 86}, 043403 (2012).
\bibitem{PETROSYAN}D. Petrosyan, M. H\"oning and M. Fleischhauer, {\em Phys. Rev. A} {\bf 87}, 053414 (2013).
\bibitem{PETROSYAN2}D. Petrosyan, Phys. Rev. A {\bf 88}, 043431 (2013).
\bibitem{PETROSYANBOOK} P. Lambropoulos and D. Petrosyan, {\em Fundamentals of Quantum Optics and Quantum Information}, Springer Berlin  (2007).
\bibitem{CHANDLER}D. Chandler, \textit{Introduction to modern statistical mechanics}, Oxford University Press (2001).
\bibitem{suppmat}For details on the origin of fluctuations in the model, the rotational invariant averaging and excitation densities for various numbers of excitations see Supplementary Material. 
\bibitem{HARDRODS}C. Ates and I. Lesanovsky, Phys. Rev. A {\bf 86}, 013408 (2012).
\bibitem{HARDRODS2}S. Ji, C. Ates, J. P. Garrahan and I. Lesanovsky, J. Stat. Mech. P02005 (2013).
\end{thebibliography}

\newpage

\clearpage

\appendix

\section{Supplementary Materials}

\subsection{Fluctuations}

The fluctuations around the equilibrium position of particles in the logarithmic potential model correspond to de-excitations of an atom succeeded by an immediate excitation of an other atom in the close vicinity (illustrated in Fig. \ref{fig:illustration}). In the following we compare the distribution of the location of these re-excitations with the thermodynamic fluctuations in the particle model. As an illustration we introduce the following simple model and analytically calculate the expected fluctuations in both models, the excitation model and the particle model. In the steady state the average distance between excitations is the blockade radius 

\begin{figure}[htb]
\centerline{\includegraphics[width=6.0cm]{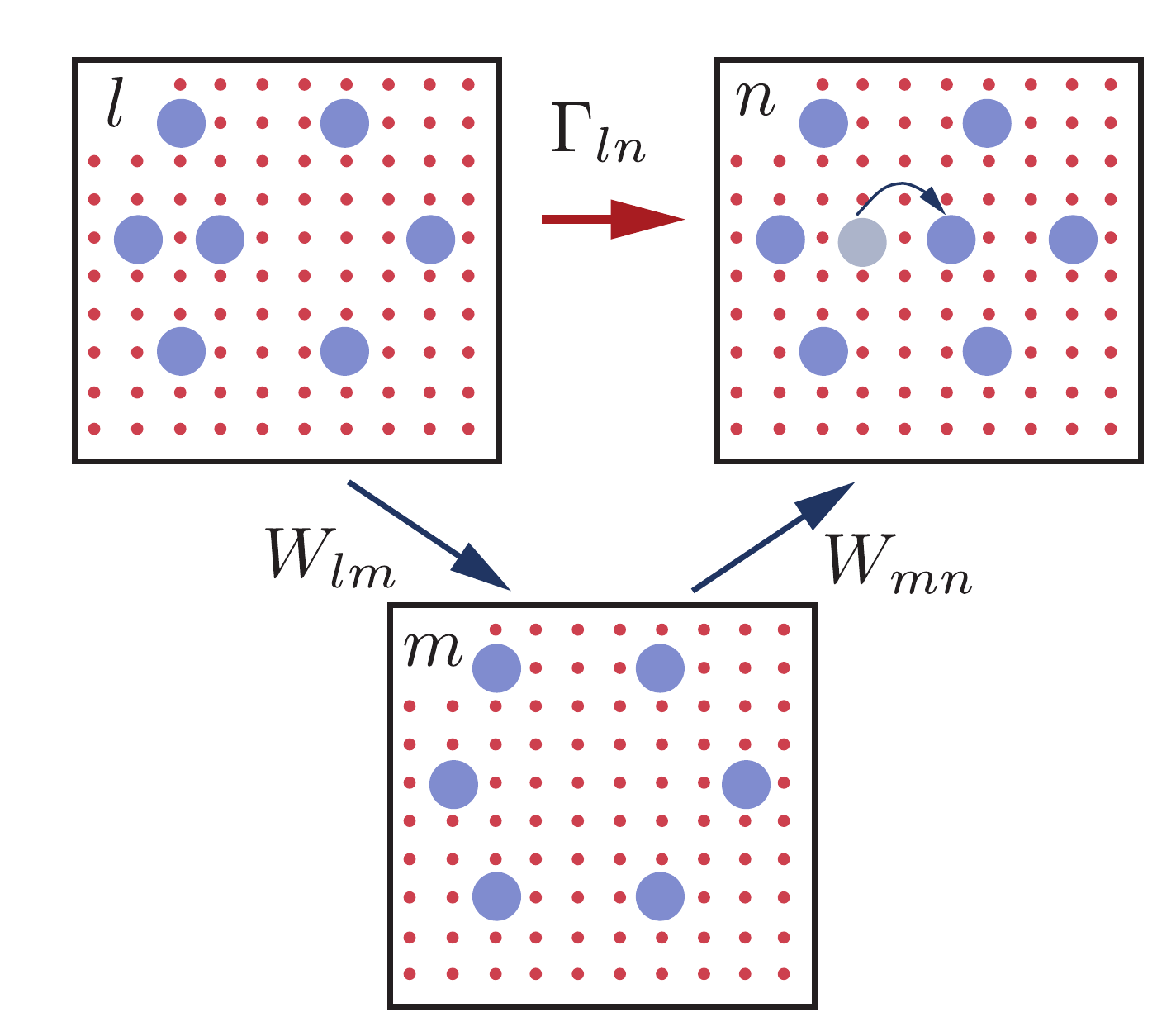}}
\caption{Illustration of the mapping: In the microscopic picture, the steady state is the result of atoms that are excited and de-excited. Consider the following sequence: In a typical steady state configuration l (top left) with $N=7$ excitations one Rydberg atom is de-excited (process $W_{lm}$). In a second step, another atom is excited at a different position (process $W_{nm}$). Assuming l was a steady state configuration, the probability is large that the atom is excited close to the position of the previously de-excited atom. In the logarithmic potential model, we consider only sub-ensembles of constant $N$, in this example $N=7$. In this interpretation, the sequence of events is a movement of an effective particle following a pair interaction.}
\label{fig:illustration}
\end{figure}

\begin{figure}[htb]
\centerline{\includegraphics[width=6.0cm]{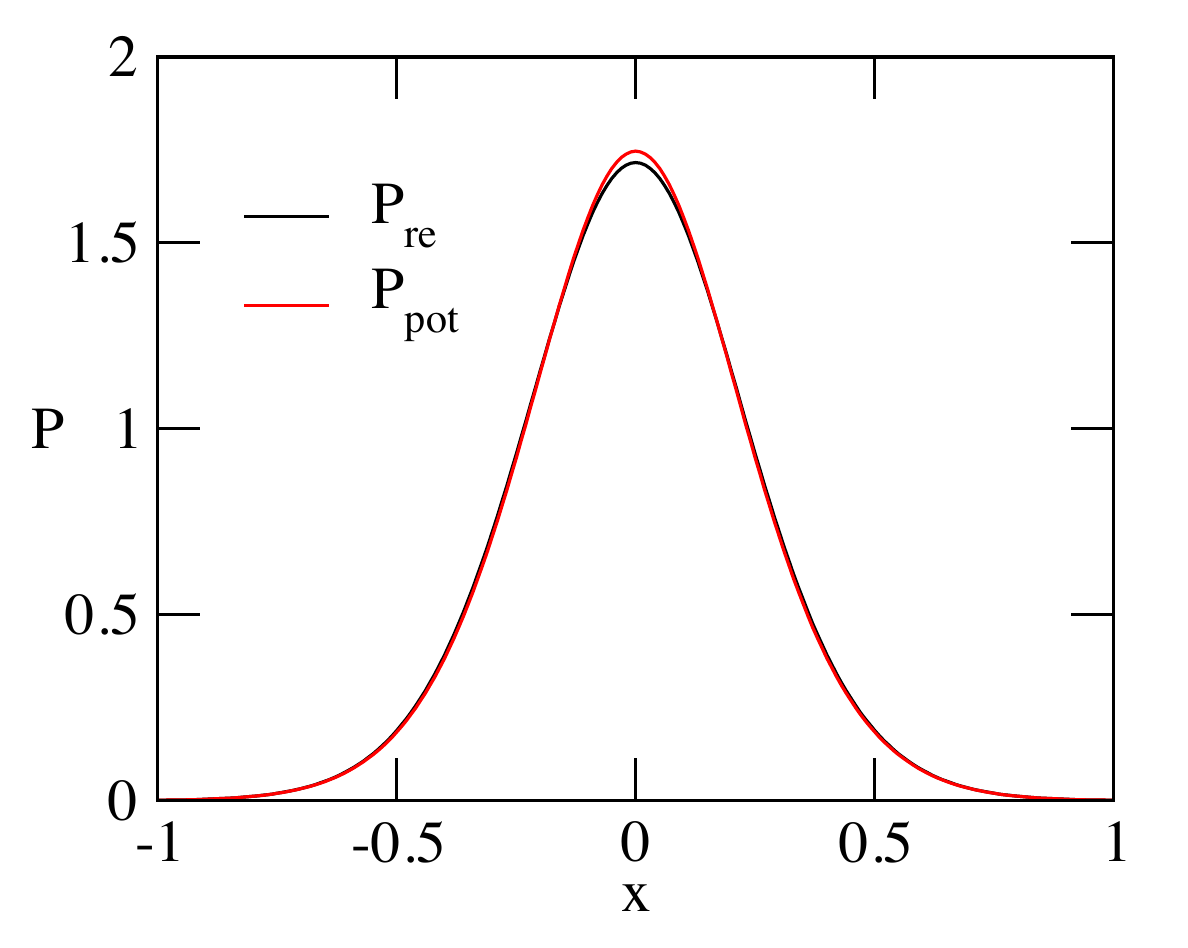}}
\caption{The distribution of the position of a re-excited atoms with two neighbors at the steady state distance compared to the fluctuations of a particle in the logarithmic potential of two neighboring particles. }
\label{fig:fluctuations}
\end{figure}

\begin{equation}
r_B = \left( \frac{C_6}{2 \Omega \sqrt{\frac{\Gamma_z}{\gamma_{rg}}}} \right)^{1/6}.
\end{equation}

We consider an excited atom at position $x=0$ with two neighbors at $x=r_B$ and $x=-r_B$, assuming only nearest neighbor interaction in a 1D system. If the atom at the origin gets de-excited it can only be re-excited in the close vicinity of the original position. The distribution of the exact position of re-excitation is then

\begin{equation}
P_{re} (x) = \frac{1}{Q_{re}}\frac{\Omega^2}{2 \Omega^2 + \frac{\Gamma}{\gamma_{rg}} [\gamma_{rg}^2  + (\frac{C_6}{(x-r_B)^6}  + \frac{C_6}{(x+r_B)^6})^2] },
\label{equ:pre}
\end{equation}
where $Q_{re} = \int_{-r_B}^{r_b} \frac{\Omega^2}{2 \Omega^2 + \frac{\Gamma}{\gamma_{rg}} [\gamma_{rg}^2  + (\frac{C_6}{(x-r_B)^6}  + \frac{C_6}{(x+r_B)^6})^2]}$ is the normalization factor. The Lindemann width of this distribution is then  

\begin{equation}
\Delta x = \frac{\sqrt{\int_{-r_B}^{r_B} P_{re} (x)^2}}{r_B}.
\end{equation}

In the particle picture, this corresponds to an atom at position $x=0$ with thermal fluctuations in the logarithmic potential from two particles fixed at $x=-r_B$ and $x=r_B$ which is 

\begin{equation}
V(x) = V_{eff}(x-r_B) + V_{eff}(x+r_B),
\end{equation}
where $V_{eff} = k_B T \Phi_2(r_{ij})$ from Eq.(5) from the main text. The equilibrium probability of the particle is 

\begin{equation}
P_{pot} = \frac{1}{Q_{pot}} e^{-\frac{V(x)}{k_B T}}.
\label{equ:ppot}
\end{equation}
Here, the normalization factor is $Q_{pot} = \int_{-r_B}^{r_b} e^{-\frac{V(x)}{k_B T}}$. 

The distributions Eq.~(S\ref{equ:ppot}) and Eq. (S\ref{equ:pre}) for typical parameters ($\Delta = 0$,$\gamma_rg = 0.1\Omega$, $\Gamma = 0.1\Omega$, $C_6 = 100 \Omega$) are in almost perfect agreement. The Lindemann parameters of these distributions are $\Delta x = 0.568$ and $\Delta x = 0.57$ for the Monte Carlo simulation and particle picture, respectively.

\subsection{Rotation Protocol for the Spherical Geometry}

The crystalline pattern in the spherical geometry in Figs 4 (a) and (b) is the result of a rotation protocol to allow comparison with Ref. [1]. We use the following method: For each excitation we calculate the distance to the center $s=\sqrt{x^2+y^2}$ and distinguish between particles at the border $s>=5$ and particles in the center $s<5$. The number of excitations on the border then serves as the symmetry the crystal is compared to (e.g if $6$ particles are on the border $N_B=6$, the crystal is compared to a hexagon). The parameter for each excitation $i$ we calculate 
\begin{equation}
\psi_{N_B} = \sum_i e^{N_B \theta_i},
\end{equation}
where $\theta_i$ is the angle of the vector from the origin to excitation $i$ with respect to the $y$ axis. The form of $\psi$ is a generalization of the hexatic order parameter $\psi_6$ which is used to classify the orientation of hexagonal plaquettes in a triangular lattice. The parameter $\psi_{N_B}$ is the relative orientation of a polygon with $N_B$ side with respect to the polygon spanned by the Rydberg excitations. Then the system is rotated to find the minimum of $\psi_{N_B}$. This rotated configuration is then added to the histogram shown in Figs. 4 (a) and (b) in the main text. Note, that for a given number of excitations several symmetries can occur. For example, for $N=6$ the crystal can either consist of $6$ excitations on the border (hexagonal) or $5$ excitations on the border (pentagonal) and $1$ excitation in the center. This is clearly seen in Fig. (4) as the pattern is a mixture of a hexagonal and a pentagonal crystal. 

\subsection{Patterns of various $N$-ensembles}

\begin{figure}[htb]
\centerline{\includegraphics[width=10.0cm]{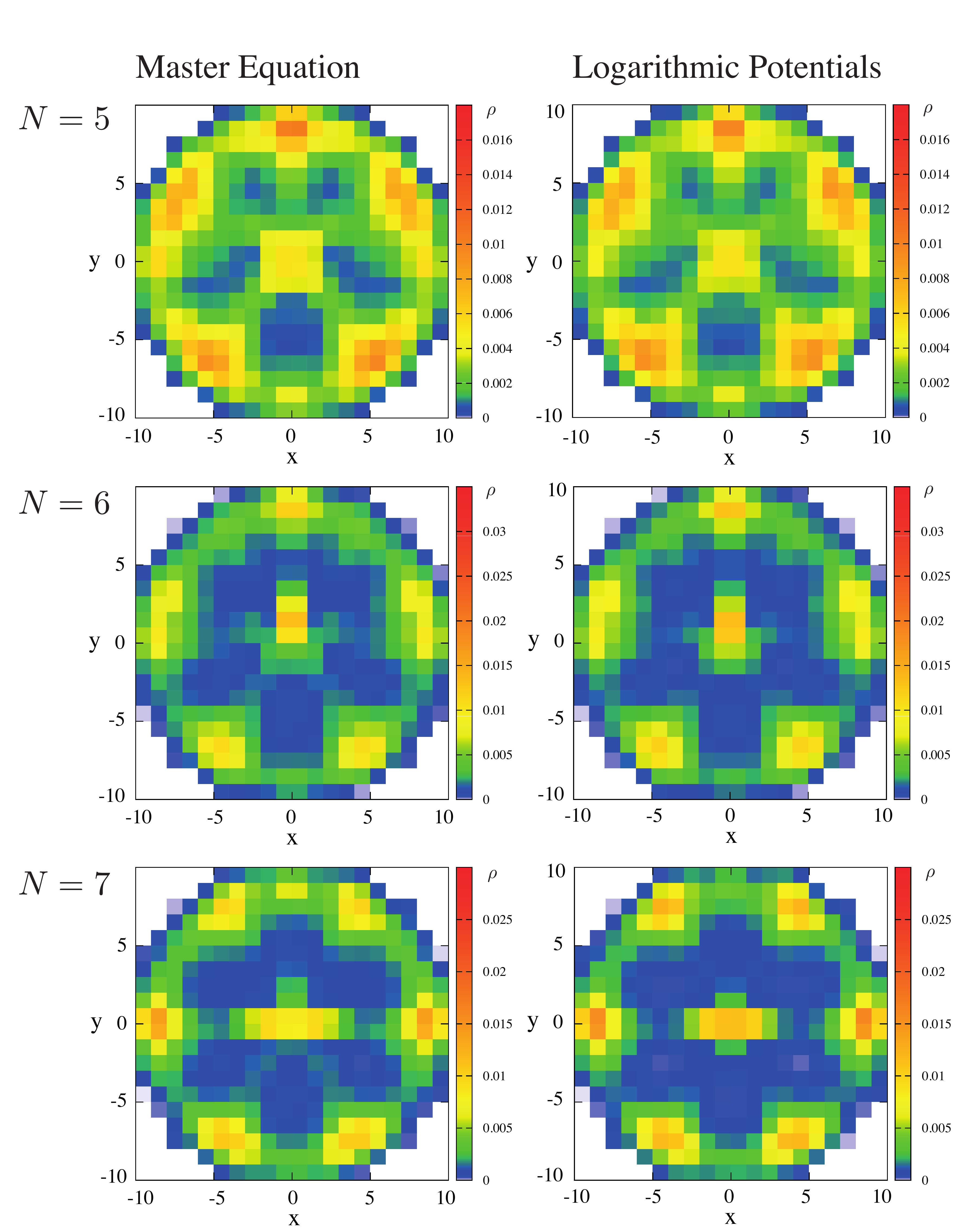}}
\caption{Series of density plots for various numbers of excitations. In the master equation approach (left), the full ensemble is split up in a post-selection step into the various $N$-manifolds. In the logarithmic potential approach (right) each $N$ corresponds to a system with $N$ particles and logarithmic pair interactions.}
\label{fig:series}
\end{figure}

The excitation patterns in Rydberg ensembles are revealed from a post-selection of configurations with $N$ excitations. This can be achieved by solving the master equation [Eq. (3) from the main text] and store configurations according to the number of excitations. The resulting patterns for $N=5,6$, and $7$ excitations is shown in Fig. \ref{fig:series} (left column). The logarithmic potential model requires an individual simulation for each number of excitation. The potential is the same in all 3 cases. The results, shown in Fig. \ref{fig:series} (right column) are in good agreement with the original model for all densities. 

\end{document}